\documentclass[pra,preprint,superscriptaddress,showpacs,amsmath,amssymb]{revtex4-1}

\usepackage{graphicx}
\usepackage[separate-uncertainty=true,multi-part-units=single,range-phrase=--,range-units=single,list-units=single]{siunitx}
\usepackage[capitalize]{cleveref}
\usepackage{microtype}
\usepackage{xcolor}
\usepackage{tensor}

\DeclareSIUnit\mJ{\milli\J}

\providecommand{\config}[3]{\smash{\ensuremath{{#1}{#2}^{#3}}}}
\providecommand{\term}[3]{\smash{\ensuremath{\tensor*[^{#1}]{\text{#2}}{_{#3}}}}}
\providecommand{\ani}[1]{\ensuremath{\text{#1}^{-}}}
\providecommand{\iso}[2]{\smash{\ensuremath{^{#1}\text{#2}}}}

\begin{document}
	
	\title{Radiative Lifetimes of the Bound Excited States of \ani{Pt}}
	\author{KC Chartkunchand}
		\email[Corresponding author: ]{kiattichart.chartkunchand@fysik.su.se}
\affiliation{Department of Physics, Stockholm University, AlbaNova, SE-106 91
Stockholm, Sweden}
	\author{M. Kami\'{n}ska}
\affiliation{Department of Physics, Stockholm University, AlbaNova, SE-106 91
Stockholm, Sweden}
\affiliation{Institute of Physics, Jan Kochanowski University, 25-369 Kielce,
Poland}
	\author{E. K. Anderson}
\affiliation{Department of Physics, Stockholm University, AlbaNova, SE-106 91
Stockholm, Sweden}
	\author{M. K. Kristiansson}
\affiliation{Department of Physics, University of Gothenburg, SE-412 96
Gothenburg, Sweden}
	\author{G. Eklund}
\affiliation{Department of Physics, Stockholm University, AlbaNova, SE-106 91
Stockholm, Sweden}
	\author{O. M. Hole}
\affiliation{Department of Physics, Stockholm University, AlbaNova, SE-106 91
Stockholm, Sweden}
	\author{R. F. Nascimento}
\affiliation{Department of Physics, Stockholm University, AlbaNova, SE-106 91
Stockholm, Sweden}
\affiliation{Centro Federal de Educa\c c\~{a}o Tecnol\'{o}gica Celso Suckow da
Fonseca, Petr\'{o}polis, 25620-003, RJ, Brazil}
	\author{M. Blom}
	\author{M. Bj\"{o}rkhage}
	\author{A. K\"{a}llberg}
	\author{P. L\"{o}fgren}
	\author{P. Reinhed}
	\author{S. Ros\'{e}n}
	\author{A. Simonsson}
	\author{R. D. Thomas}
	\author{S. Mannervik}
\affiliation{Department of Physics, Stockholm University, AlbaNova, SE-106 91
Stockholm, Sweden}
	\author{V. T. Davis}
	\author{P. A. Neill}
	\author{J. S. Thompson}
\affiliation{Department of Physics, University of Nevada, Reno, Nevada 89557,
United States}
	\author{D. Hanstorp}
\affiliation{Department of Physics, University of Gothenburg, SE-412 96
Gothenburg, Sweden}
	\author{H. Zettergren }
	\author{H. Cederquist}
	\author{H. T. Schmidt}
		\email{schmidt@fysik.su.se}
\affiliation{Department of Physics, Stockholm University, AlbaNova, SE-106 91
Stockholm, Sweden}
	
	\begin{abstract}
The intrinsic radiative lifetimes of the \config{5}{d}{10}\config{6}{s}{}
\term{2}{S}{1/2} and \config{5}{d}{9}\config{6}{s}{2} \term{2}{D}{3/2} bound
excited states in the platinum anion \ani{Pt} have been studied at cryogenic
temperatures at the DESIREE (Double ElectroStatic Ion Ring Experiment) facility
at Stockholm University. The intrinsic lifetime of the higher-lying
\config{5}{d}{10}\config{6}{s}{} \term{2}{S}{1/2} state was measured to be
\SI{2.54(10)}{\s}, while only a lifetime in the range of \SIrange{50}{200}{\ms}
could be estimated for the \config{5}{d}{9}\config{6}{s}{2} \term{2}{D}{3/2}
fine-structure level. The storage lifetime of the \ani{Pt} ion beam was measured
to be a little over 15 minutes at a ring temperature of \SI{13}{\K}. The present study
is the first to report the lifetime of an atomic negative ion in an excited
bound state with an electron configuration different from that of the ground
state.
	\end{abstract}
	
	
	\date{\today}
	
	\maketitle
	
	\section{Introduction}
Negative ions are fascinating quantum systems that illustrate the subtle nature
of electronic structure and dynamics on the atomic scale. Due to the short range
of the forces that hold these systems together, the number of bound quantum
states is strongly limited. Some atoms do not even form stable anions, or have
only a single bound state~\cite{Peterson85,Andersen04}. Other atomic anions,
however, possess one or a few bound \emph{excited} states. These include
higher-lying $LS$ terms or fine-structure levels of the same electron
configuration as the anionic ground state~\cite{Hotop85,Pegg04} and/or bound
excited states with different electron configurations~\cite{Scheer98}. In most
cases, the ground and excited atomic anionic states have the same parity.
Single-photon electric dipole (E1) transitions are then forbidden by the parity
selection rule and the corresponding excited states are often very long-lived.
This makes studies based on conventional optical spectroscopy methods difficult
and the lifetimes of bound excited states in atomic anions are, with few
exceptions, unknown. From a theoretical perspective, studies of negative ions
are made more challenging by the need to accurately describe electron
correlation effects, which are much more important for anions than for neutrals
or cations. Measurements of negative ion properties such as binding energies and
lifetimes can thus provide critical benchmarks for theoretical methods treating
such correlation effects in atomic systems in general.

The characterization of atomic and molecular anions is crucial for several
fields of science, as well as for certain technological applications. Negative
ions are, for example, believed to play key roles as absorbers, emitters, and
reaction partners in stellar and planetary atmospheres, and in the interstellar
medium~\cite{Goudsmit34,Wildt37,Wildt44,Massey40,Warner67,Strittmatter71,Sarre00}.
Additionally, the ability to produce beams of atomic anions forms the basis for
a range of powerful and widely-used radiological dating
techniques~\cite{Fifield99,Pegg04}. Furthermore, the discovery of bound excited
states in \ani{La}~\cite{Walter14}, \ani{Ce}~\cite{Walter11}, and
\ani{Os}~\cite{Bilodeau00,Kellerbauer14} with parities opposite to those of the
corresponding ground states has opened up the possibility of laser cooling of
these atomic anions. Such cold anions could be used to sympathetically cool
anti-protons to facilitate the production of ultra-cold anti-hydrogen for use in
high-precision spectroscopic tests of CPT symmetry, and for measurements of the
effects of gravity on anti-matter~\cite{Kellerbauer06}. In each of the above
examples, a detailed understanding of anion properties is absolutely necessary.

Only recently have techniques for long-term storage of negative ions (seconds,
minutes, and beyond) become available, allowing for studies of long-lived
anionic excited bound
states~\cite{Balling92,Haugen92,Andersen93,Moller97,Zajfman97,Schmidt01,Trabert02,Ellmann04,Andersson06,Bernard08,Thomas11,vonHahn11,Nakano12}.
The cryogenic ion storage rings DESIREE~\cite{Thomas11,Schmidt13,Backstrom15},
CSR~\cite{vonHahn11}, and RICE~\cite{Enomoto15}, which can be used to store
\si{\keV} ion beams for up to hours in cryogenic, magnetic-field-free
environments, are at the forefront in this field~\cite{Schmidt15}. Some of the
capabilities of the DESIREE facility were recently demonstrated through
measurements of the radiative lifetimes of the long-lived, excited
fine-structure level in the \config{n}{p}{5} \term{2}{P}{1/2} $\rightarrow$
\config{n}{p}{5} \term{2}{P}{3/2} transition in \ani{S} ($n=3$,
$\tau=\SI{503(54)}{\s}$~\cite{Backstrom15}), \ani{Se} ($n=4$,
$\tau=\SI{4.78(18)}{\s}$~\cite{Backstrom15}), and \ani{Te} ($n=5$,
$\tau=\SI{0.463(8)}{\s}$~\cite{Backstrom15}), and of the
\config{3}{d}{9}\config{4}{s}{2} \term{2}{D}{3/2} $\rightarrow$
\config{3}{d}{9}\config{4}{s}{2} \term{2}{D}{5/2} transition in \ani{Ni}
($\tau=\SI{15.1(4)}{\s}$)~\cite{Kaminska16}.
	
	\begin{figure}[htp]
		\centering
		\includegraphics[width=\columnwidth]{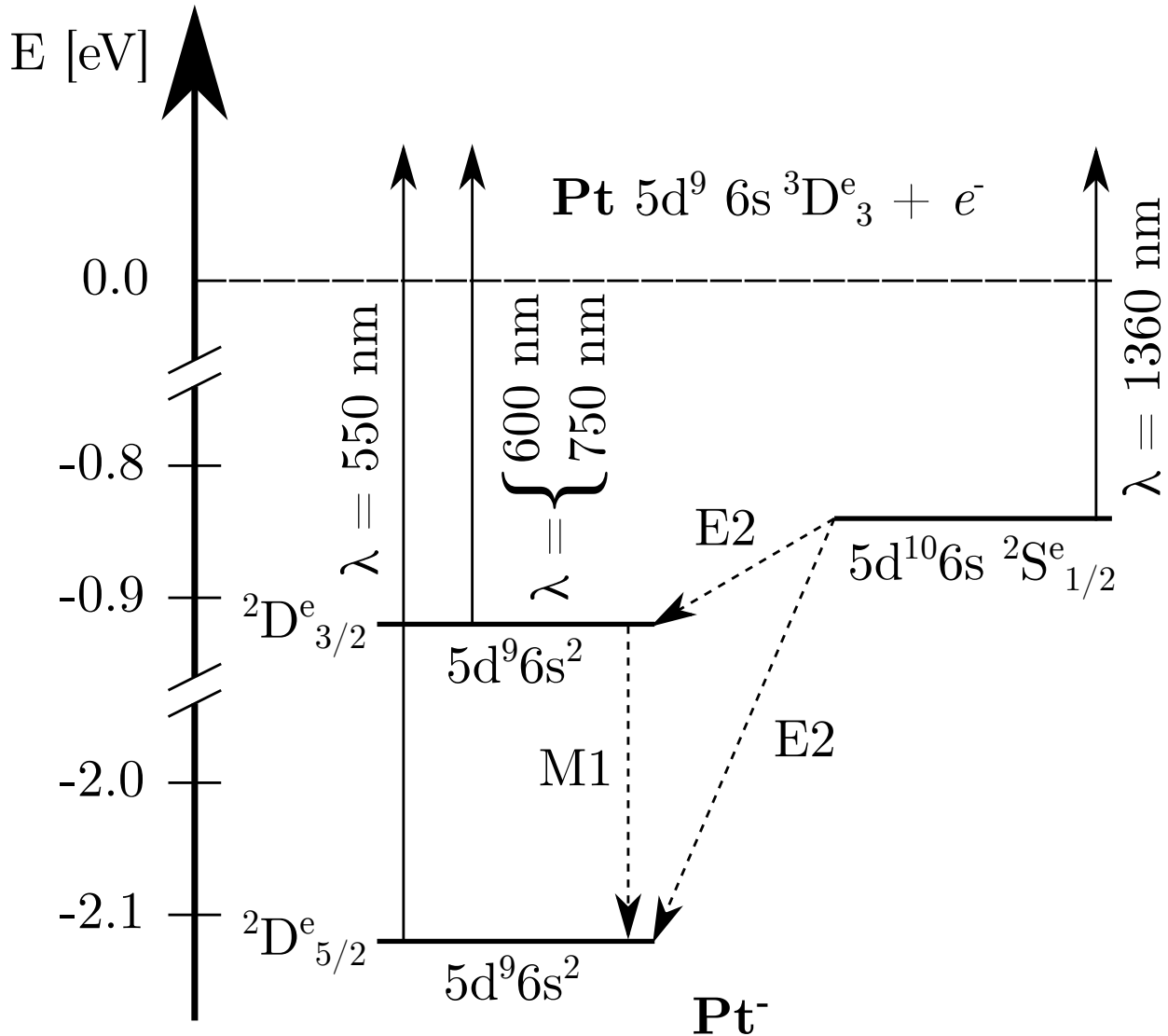}
\caption{Energy-level diagram of the three bound states of \ani{Pt}. The
vertical scale indicates total energies with respect to the ground state of the
neutral Pt atom. Magnetic dipole (M1) and electric quadrupole (E2) transitions
are the lowest-order transitions possible between the ground and excited states
of \ani{Pt}.}
		\label{PtED}
	\end{figure}
The electronic structure of the platinum anion, \ani{Pt}, has been studied
previously and its energy level diagram is shown in \cref{PtED}. The electron
affinity of Pt was measured to be
\SI[separate-uncertainty=false]{2.12510(5)}{\eV} by means of the laser
photodetachment threshold (LPT) technique~\cite{Bilodeau99}. Two-photon
detachment spectroscopy was utilized to determine the excitation energy of the
\config{5}{d}{9}\config{6}{s}{2} \term{2}{D}{3/2} fine-structure level, which
was found to lie \SI[separate-uncertainty=false]{1.20775(6)}{\eV} above the
\config{5}{d}{9}\config{6}{s}{2} \term{2}{D}{5/2} anionic ground
state~\cite{Thogersen96}. The highest-lying \config{5}{d}{10}\config{6}{s}{}
\term{2}{S}{1/2} \ani{Pt} bound state, which has a different electron
configuration from that of the ground state, was measured to lie
\SI[separate-uncertainty=false]{1.27567(161)}{\eV} above the \ani{Pt} ground
state~\cite{Andersson09}. Both excited states are expected to be long-lived
since E1 transitions to the ground state are forbidden by the parity selection
rule. The \config{5}{d}{9}\config{6}{s}{2} \term{2}{D}{3/2} fine-structure level
may decay by an M1 transition to the ground state, while decay from the
\config{5}{d}{10}\config{6}{s}{} \term{2}{S}{1/2} state may proceed through E2
transitions to the \config{5}{d}{9}\config{6}{s}{2} \term{2}{D}{5/2,3/2} levels.
		
Th{\o}gersen \emph{et al}.~\cite{Thogersen96} used multi-configurational
Dirac-Fock methods to calculate a spontaneous decay rate of
$\Gamma=\SI{14}{\per\s}$ for the \config{5}{d}{9}\config{6}{s}{2}
\term{2}{D}{3/2} $\rightarrow$ \config{5}{d}{9}\config{6}{s}{2} \term{2}{D}{5/2}
transition in \ani{Pt}, corresponding to a lifetime of $\tau=\SI{71}{\ms}$. It
is interesting to note that a scaling of the $\tau=\SI{15.1(4)}{\s}$
experimental lifetime of the \config{3}{d}{9}\config{4}{s}{2} \term{2}{D}{3/2}
level in \ani{Ni}~\cite{Kaminska16} with the third power of the ratio between
the corresponding transition energies in \ani{Ni} and \ani{Pt} suggests a
lifetime of the \config{5}{d}{9}\config{6}{s}{2} \term{2}{D}{3/2} level in
\ani{Pt} of $\tau=\SI{53(2)}{\ms}$~\cite{BrageARXIV}. To the best of our
knowledge, no calculation of the \config{5}{d}{10}\config{6}{s}{}
\term{2}{S}{1/2} $\rightarrow$ \config{5}{d}{9}\config{6}{s}{2}
\term{2}{D}{3/2,5/2} transition rates in \ani{Pt}, which are the main quantities
of interest in the present study, has been reported in the literature.

	\section{Experimental Apparatus}
	\begin{figure*}[htp]
		\centering
		\includegraphics[width=\linewidth]{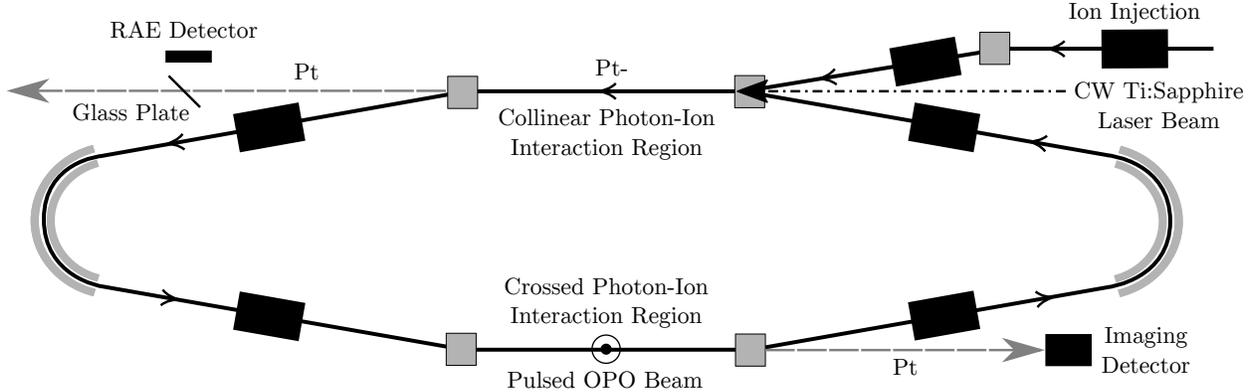}
\caption{Schematic of one of the ion storage rings of DESIREE. \SI{10}{\keV}
beams of \ani{Pt} are injected and stored in the ring. The output of a
continuous wave (CW) Ti:Sapphire laser propagates along the ion beam direction
in one of the straight sections of the ring and can be overlapped with the
stored ions in a collinear geometry, while the output of a pulsed optical
parametric oscillator (OPO) can be crossed with the ion beam at \ang{90} in the
other straight section. Neutral Pt atoms produced through photodetachment in the
collinear interaction region are detected by a resistive anode encoder (RAE)
detector, while those produced in the crossed interaction region are detected by
an imaging detector.}
		\label{DESIREE}
	\end{figure*}
\begin{figure}[htp]
		\centering
		\includegraphics[width=\columnwidth]{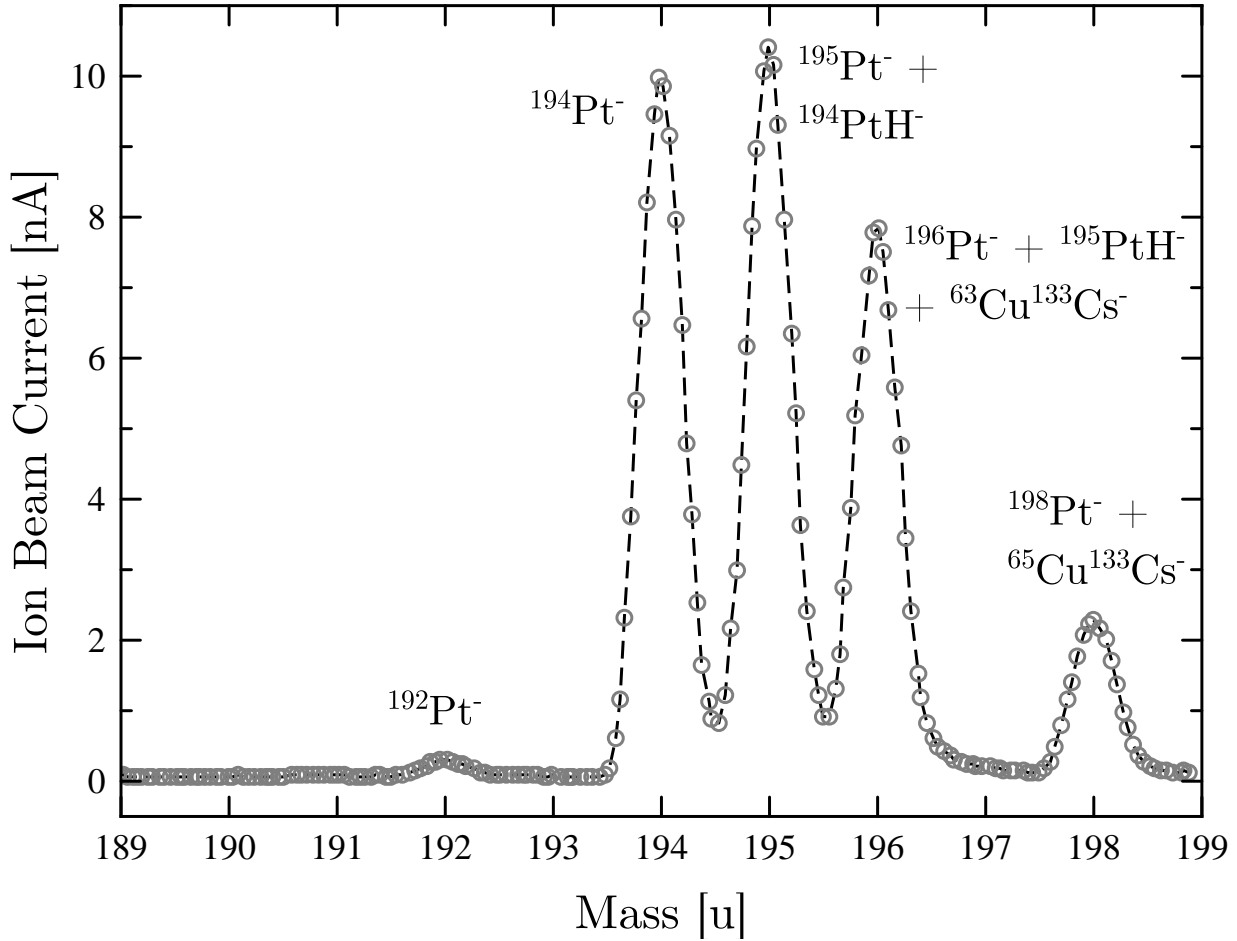}
\caption{Mass scan of \ani{Pt} ion beam in the range of
\SIrange{190}{199}{\amu}. The \iso{194}{\ani{Pt}} isotope was used throughout
the experiment to avoid contamination from other anionic species that share the
same mass as the \SIlist{195;196;198}{\amu} isotopes of Pt.}
		\label{PtMS}
	\end{figure}

The present experiment was conducted at the cryogenic electrostatic ion beam
storage ring DESIREE. A full description of this facility is provided in
Refs.~\cite{Thomas11,Schmidt13}, so only a brief summary is given here. Platinum
anions (\ani{Pt}) were produced in a SNICS II cesium-sputter ion
source~\cite{SNICS}. The \ani{Pt} ions were accelerated to an energy of
\SI{10}{\keV} and passed through a \ang{90} analyzing magnet for mass selection
before injection into the symmetric ion storage ring of
DESIREE~\cite{Thomas11,Schmidt13}, as shown in \cref{DESIREE}. A typical mass
scan of the \ani{Pt} beam is shown in \cref{PtMS}. The ratios of ion beam
currents produced for each \ani{Pt} mass agreed well with the natural abundances
of Pt isotopes, indicating that only small amounts of \ani{PtH} were produced in
the ion source. For the present experiment, the \iso{194}{\ani{Pt}} isotope was
selected to minimize the risk of contamination by other anionic species at
masses \SIlist{195;196;198}{\amu}. Such contaminations were observed in earlier
experiments utilizing a similar type of ion source~\cite{Andersson09}. As
examples, the \iso{194,195}{Pt}\ani{H} and \iso{63,65}{Cu}\iso{133}{Cs}$^{-}$
molecular anions overlap some \ani{Pt} masses above \SI{194}{\amu}.
	
Two different photon sources and interaction geometries were used for probing
the populations of the three bound states of the \ani{Pt} ions as functions of
time after injection in the ion storage ring. A continuous wave (CW) Sirah
Matisse Ti:Sapphire laser was used in a collinear interaction geometry in one of
the straight sections of the ion storage ring as shown in~\cref{DESIREE}. The CW
laser beam was propagated along the ion beam direction, allowing for a large
photon-ion interaction volume. Neutral Pt atoms produced by photodetachment
leave the ring and strike a gold-coated glass plate, producing secondary
electrons that are registered by a particle detector consisting of three
microchannel plates (MCPs) and a resistive anode encoder
(RAE)~\cite{Rosen07,Thomas11,Schmidt13}. Alternatively, the output of an Ekspla
NT300-series tunable pulsed optical parametric oscillator (OPO), with a pulse
width of $\sim$\SI{5}{\ns} and repetition rate of \SI{10}{\Hz}, was used in a
crossed-beam interaction geometry in the opposite straight section of the ring
(see~\cref{DESIREE}). In this case, neutral Pt atoms are counted by an imaging
detector, consisting of a triple-stack MCP with a phosphor screen anode viewed
by a photomultiplier tube~\cite{Rosen07,Thomas11,Schmidt13}. In both cases,
background signals were mainly due to detector dark counts and only to a limited
extent to neutral Pt atoms produced through collisions with the very dilute
residual gas of \numrange{e3}{e4} H$_{2}$ molecules per
\si{\cubic\cm}~\cite{Schmidt13}.
	
The lifetimes of the bound states of \ani{Pt} were measured by state-selective
probing of their populations as functions of time after ion injection using
either the OPO or the Ti:Sapphire laser. The pulsed OPO offered a broader
wavelength tunability (\SIrange{200}{2600}{\nm}) compared to the Ti:Sapphire
laser (\SIrange{700}{990}{\nm}), allowing for detachment from only the
\config{5}{d}{10}\config{6}{s}{} \term{2}{S}{1/2} excited state with
\SI{1360}{\nm} light (\SI{0.912}{\eV} photon energy), detachment from both
excited states with \SI{600}{\nm} light (\SI{2.07}{\eV} photon energy), and
detachment from all three bound states with \SI{550}{\nm} light (\SI{2.25}{\eV}
photon energy). The short pulse length of the OPO also allowed for efficient
background suppression by only counting Pt atoms on the imaging detector within
a \SI{10}{\micro\s}-wide time window around the OPO pulse. The OPO pulse energy
and ion beam current were adjusted such that the probability for a given OPO
pulse to produce one Pt atom was $<0.01$. This was done in order to limit the
probability that two neutrals were produced by a single OPO pulse, which would
be counted as only one event by the detector due to its finite response time
(resulting in a so-called ``counting saturation''). Background contributions
from collisions in the residual gas and detector dark counts were measured by
counting events within a \SI{40}{\ms}-wide time window between the OPO pulses.

The CW Ti:Sapphire laser could produce a higher time-averaged neutral signal
rate via photodetachment compared to the pulsed OPO. This was due to a
combination of the much larger ion-photon interaction volume provided by the
collinear interaction geometry and the use of higher time-averaged photon
intensities and ion beam currents since ``counting saturation'' was no longer an
issue with the CW laser. The CW mode of the Ti:Sapphire laser was also better
suited for studying lifetimes shorter than \SI{100}{\ms}, a situation in which
the fixed \SI{10}{\Hz} repetition rate of the OPO became a limiting factor. The
Ti:Sapphire laser was used at a wavelength of \SI{750}{\nm} (\SI{1.65}{\eV}
photon energy), allowing for photodetachment from both excited states of
\ani{Pt}.
	
	\section{Results and Discussion}
In order to extract effective lifetimes from the measured decay curves, fitting
functions in the form of exponentials or sums of exponentials were
used~\cite{Kaminska16}. A weighted, non-linear least-squares routine was used in
fitting the data, with each data point weighted according to its statistical
uncertainty assuming a Poisson distribution. Each fit resulted in the effective
lifetime of each \ani{Pt} state, denoted by $\tau^{\text{eff}}_{5/2}$,
$\tau^{\text{eff}}_{3/2}$, and $\tau^{\text{eff}}_{1/2}$ for the
\config{5}{d}{9}\config{6}{s}{2} \term{2}{D}{5/2},
\config{5}{d}{9}\config{6}{s}{2} \term{2}{D}{3/2}, and
\config{5}{d}{10}\config{6}{s}{} \term{2}{S}{1/2} states, respectively. The
effective lifetime of the anionic ground state, $\tau^{\text{eff}}_{5/2}$,
corresponds to the storage lifetime of the \ani{Pt} ions in the ring. Photon
intensities used to probe the ions were adjusted such that photodetachment rates
remained negligible in relation to the effective decay rates to be measured. The
effective decay rate of a bound excited state in the stored ion beam is then the
sum of the corresponding intrinsic decay rate and the rate at which the excited
ions are lost through collisional detachment with the residual H$_{2}$ gas or
through ion storage imperfections (e.g., voltage ripples on the ion optical
elements in the ring)~\cite{Schmidt13,Backstrom15}.

\subsection{The \config{5}{d}{10}\config{6}{s}{} \term{2}{S}{1/2} $\rightarrow$
\config{5}{d}{9}\config{6}{s}{2} \term{2}{D}{5/2} Transition}
	\begin{figure}[htp]
		\centering
		\includegraphics[width=\columnwidth]{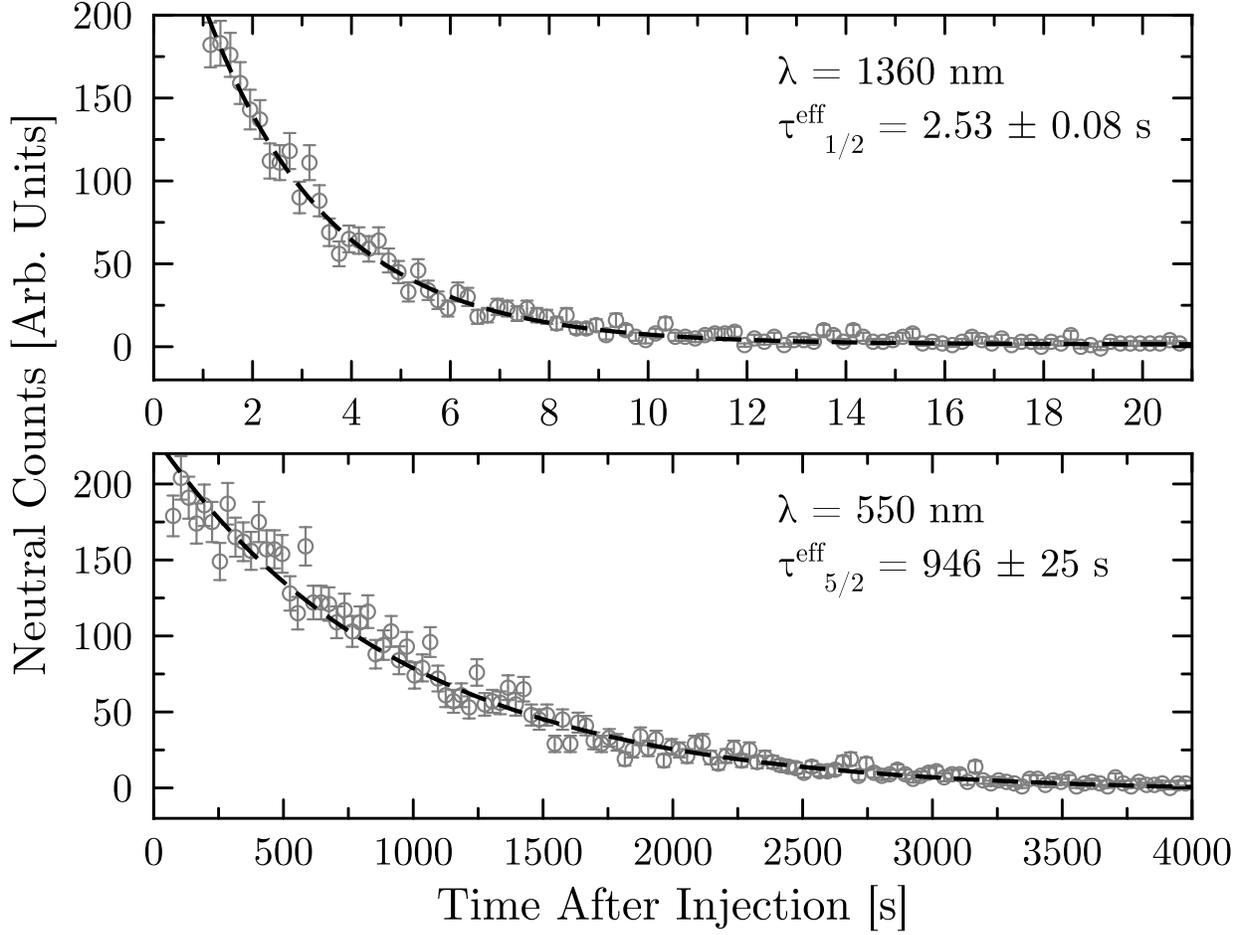}
\caption{Decay curves of the \ani{Pt} beam as a function of time at the base
storage ring temperature of \SI{13}{\K}. The upper plot shows decay signal from
only the \config{5}{d}{10}\config{6}{s}{} \term{2}{S}{1/2} state, while the
lower plot contains contributions from all three states of \ani{Pt}. Note the
difference in time scales between the upper and lower plots.}
		\label{DecayCurves}
	\end{figure}
The measured decay curve for the \config{5}{d}{10}\config{6}{s}{}
\term{2}{S}{1/2} state is shown in the upper plot of~\cref{DecayCurves}, along
with the decay curve from which the ion beam storage time is determined (lower
plot). The effective lifetime for the \config{5}{d}{10}\config{6}{s}{}
\term{2}{S}{1/2} state was found to be
$\tau^{\text{eff}}_{1/2}=\SI{2.53(08)}{\s}$, while the effective ion storage
lifetime was found to be $\tau^{\text{eff}}_{5/2}=\SI{946(25)}{\s}$. The latter
lifetime yields an ion beam decay rate of
$\Gamma^{\text{eff}}_{5/2}=1/\tau^{\text{eff}}_{5/2}=\SI{0.00106(3)}{\per\s}$.
This value can be used to deduce the \emph{intrinsic} lifetime
$\tau^{\text{int}}_{1/2}$ of the excited \term{2}{S}{1/2} state from the
corresponding \emph{effective} lifetime. The intrinsic lifetime is determined by
applying a small correction to $\tau^{\text{eff}}_{1/2}$ that accounts for the
fact that \ani{Pt} ions in the excited state may be lost in collisions with the
residual gas or by imperfections in the ion storage condition. This is expressed
by
$1/\tau^{\text{int}}_{1/2}=1/\tau^{\text{eff}}_{1/2}-[\beta+\eta(1-\beta)]\times\Gamma^{\text{eff}}_{5/2}$,
in which $\beta$ is the fraction of losses due to storage imperfections and
$\eta$ is the ratio between the cross-sections for collisional detachment in
H$_{2}$ (which completely dominates the residual gas at
\SI{13}{\K}~\cite{Schmidt13}) of \ani{Pt} in the excited and ground states. By
assuming $\beta+\eta(1-\beta)=1\pm10$, an intrinsic lifetime for the
\config{5}{d}{10}\config{6}{s}{} \term{2}{S}{1/2} state of
$\tau^{\text{int}}_{1/2}=\SI{2.54(10)}{\s}$ is determined. This is an extremely
conservative assumption; in other cases, $\eta$ has been measured to be close to
one~\cite{Backstrom15,Kaminska16}.
	
The \term{2}{S}{1/2} state is expected to decay to the anionic ground state and
the \config{5}{d}{9}\config{6}{s}{2} \term{2}{D}{3/2} excited fine-structure
level via E2 transitions (see~\cref{PtED}). An estimate of the ratio between the
E2 decay rates can be made by noting that the E2 spontaneous transition rate is
proportional to the square of the matrix element of the electric quadrupole
operator between the \term{2}{S}{1/2} and \term{2}{D}{5/2,3/2} states and the
fifth power of the energy separation between the \term{2}{S}{} and \term{2}{D}{}
states (see, e.g.,~\cite{Corney77}). Assuming that the matrix elements between
the \term{2}{S}{1/2} state and the two \config{5}{d}{9}\config{6}{s}{2}
\term{2}{D}{5/2,3/2} states are of a similar magnitude, and using the
experimentally-determined energy separations~\cite{Thogersen96,Andersson09}, the
transition to the \term{2}{D}{3/2} level can be estimated to be seven
orders-of-magnitude weaker than the transition to the \term{2}{D}{5/2} ground
state. The \config{5}{d}{10}\config{6}{s}{} \term{2}{S}{1/2} $\rightarrow$
\config{5}{d}{9}\config{6}{s}{2} \term{2}{D}{3/2} decay rate is thus estimated
to be in the \SI{e-8}{\per\s} range by using the \SI{2.54(10)}{\s} lifetime of
the upper state.
	
\subsection{The \config{5}{d}{9}\config{6}{s}{2} \term{2}{D}{3/2} $\rightarrow$
\config{5}{d}{9}\config{6}{s}{2} \term{2}{D}{5/2} Transition}
For the \config{5}{d}{9}\config{6}{s}{2} \term{2}{D}{3/2} excited fine-structure
level of \ani{Pt}, exploratory measurements were first carried out at an OPO
wavelength of \SI{600}{\nm} over a \SI{20}{\s} measurement window, i.e., the
populations in this state \emph{and} in the higher-lying \term{2}{S}{1/2} state
were probed for times up to \SI{20}{\s} after injection of ions into the ring.
Analysis of the measured decay curve revealed a decay time on the order of the
effective lifetime of the \config{5}{d}{10}\config{6}{s}{} \term{2}{S}{1/2}
excited state, with no indication of any other components that may be due to
decay from the \config{5}{d}{9}\config{6}{s}{2} \term{2}{D}{3/2} level. There
are two possibilities regarding this state: either its effective lifetime is
very similar to that of the higher-lying \config{5}{d}{10}\config{6}{s}{}
\term{2}{S}{1/2} state, or it decays on a much shorter time scale and is thus
more difficult to detect due to the ``background'' from the decay of the
higher-lying state. In light of the theoretical prediction for the lifetime of
the \config{5}{d}{9}\config{6}{s}{2} \term{2}{D}{3/2} level of
\SI{71}{\ms}~\cite{Thogersen96}, the possibility for a short \term{2}{D}{3/2}
lifetime was investigated using the CW Ti:Sapphire laser at a wavelength of
\SI{750}{\nm} in the collinear interaction geometry (see~\cref{DESIREE}).
	\begin{figure}[htp]
		\centering
		\includegraphics[width=\columnwidth]{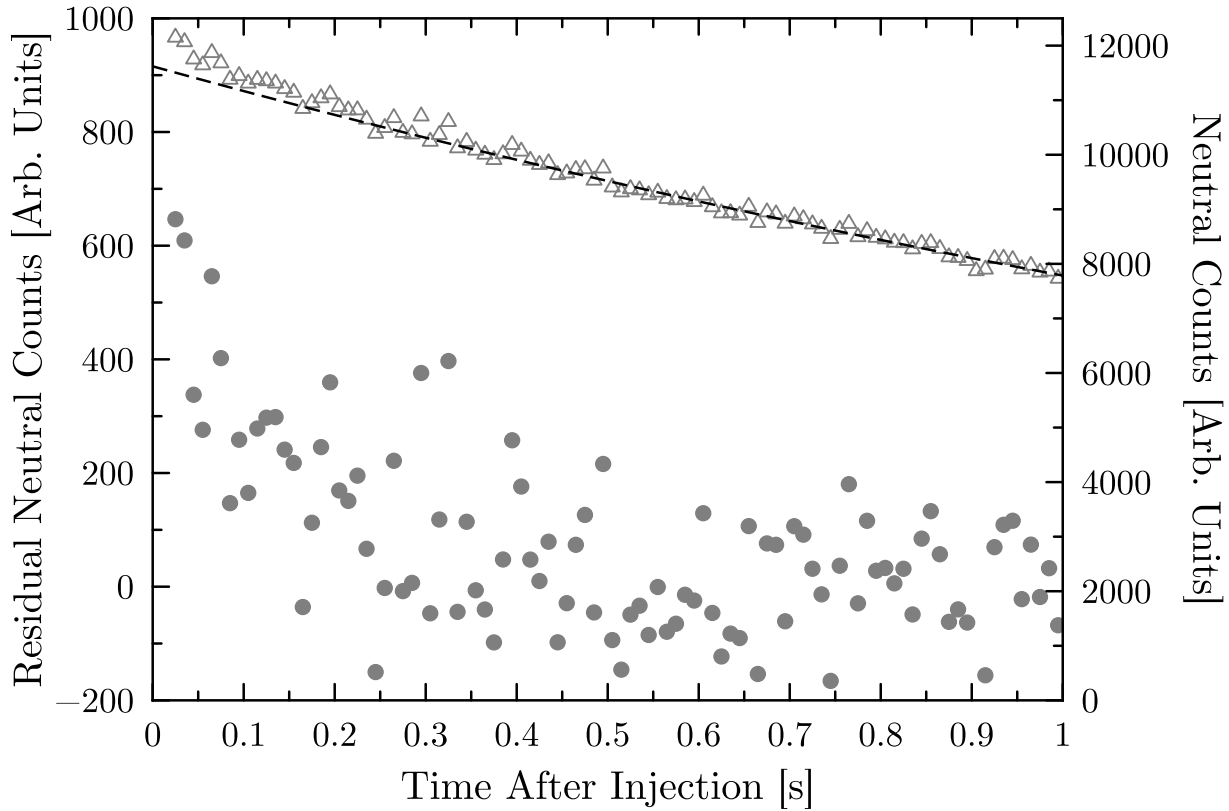}
\caption{Decay curve and residual signal of the \ani{Pt} beam as a function of
time at a Ti:Sapphire wavelength of \SI{750}{\nm}. The right vertical axis shows
the original neutral counts of the decay curve ($\bigtriangleup$) along with the
exponential function fitted to the \SI{15}{\s} measurement window data (dashed
line), while the left vertical axis shows the residual neutral counts after
subtraction of the long-decay component corresponding to decay from the
\config{5}{d}{10}\config{6}{s}{} \term{2}{S}{1/2} excited state
($\textcolor{gray}{\bullet}$).}
		\label{ResCount}
	\end{figure}	
Measurements of decay curves were performed over \SI{1}{\s} and \SI{15}{\s}
measurement windows, using $\sim$\SI{3}{\nA} of \ani{Pt} ion beam current in
both cases. The measurement with the \SI{15}{\s} time window revealed an
effective lifetime for the excited states around \SI{2.6}{\s}, which is in
agreement with the effective lifetime of the \config{5}{d}{10}\config{6}{s}{}
\term{2}{S}{1/2} state as measured with the OPO. Results from the measurement
with the \SI{1}{\s} time window are shown in~\cref{ResCount}. In order to test
for the presence of any short-lived component in the decay curve, the
exponential function used to fit the \SI{15}{\s} measurement window data was
subtracted from the \SI{1}{\s} data, thereby removing contributions from the
longer-lived \term{2}{S}{1/2} excited state. This procedure revealed the
presence of a short-lived signal which disappears $\sim$\SI{100}{\ms} after
injection, which may be due to decay of the \config{5}{d}{9}\config{6}{s}{2}
\term{2}{D}{3/2} excited state. Attempts at a more direct measurement of the
short-lived signal by increasing the ion beam current ($>$\SI{10}{\nA}) to yield
a larger overall signal rate were not fully conclusive. This was attributed to
various ion-ion effects such as intra-beam scattering, whose relaxation times
could be similar to the times in which decay from the
\config{5}{d}{9}\config{6}{s}{2} \term{2}{D}{3/2} state would be observed. Due
to these difficulties, only a range of \SIrange{50}{200}{\ms} could be estimated
for the lifetime of the \config{5}{d}{9}\config{6}{s}{2} \term{2}{D}{3/2} bound
excited state.
	
	\section{Summary and Conclusions}
In conclusion, the lifetimes of the two excited states of the \ani{Pt} anion are
reported. The highest-lying bound excited state in \ani{Pt},
\config{5}{d}{10}\config{6}{s}{} \term{2}{S}{1/2}, was found to have an
intrinsic lifetime of \SI{2.54(10)}{\s}, while only a range of
\SIrange{50}{200}{\ms} could be estimated for the intrinsic lifetime of the
\config{5}{d}{9}\config{6}{s}{2} \term{2}{D}{3/2} excited fine-structure level.
This lifetime range is consistent with available theoretical predictions for the
\term{2}{D}{3/2} state lifetime~\cite{Thogersen96,BrageARXIV}. Based on
estimates of the ratio of E2 decay rates from the \term{2}{S}{1/2} state, the
estimated lifetime of the \term{2}{S}{1/2} state via the
\config{5}{d}{10}\config{6}{s}{} \term{2}{S}{1/2} $\rightarrow$
\config{5}{d}{9}\config{6}{s}{2} \term{2}{D}{3/2} E2 transition is on the order
of 100 days. Therefore, the measured lifetime of the
\config{5}{d}{10}\config{6}{s}{} \term{2}{S}{1/2} excited state directly yields
the \config{5}{d}{10}\config{6}{s}{}
\term{2}{S}{1/2}$\rightarrow$\config{5}{d}{9}\config{6}{s}{2} \term{2}{D}{5/2}
transition rate.
	
To date, this is the first measurement of a lifetime of a bound excited state of
an atomic negative ion with a \emph{different} electron configuration than that
of the ground state. The present results may thus be used to benchmark \emph{ab
initio} calculations that include electron correlation effects, which can be
expected to differ significantly between bound anionic states of different
electron configurations. This is in contrast to the situation involving
transitions between fine-structure levels of the same $LS$ term and electron
configuration, in which the M1 transition line strength is independent of the
radial part of the wavefunction and only depends on the $L$, $S$, and $J$
quantum numbers within the $LS$-coupling approximation (a good approximation for
lighter negative ions).
	
Refinement of the experimental methods used in this study can be fruitfully
applied to studies of other atomic negative ions with bound excited states. This
includes the palladium anion \ani{Pd}, which is the final stable Group 10
negative ion yet to be studied in terms of excited state lifetimes, and the
Lanthanide negative ions, several of which have been predicted to have bound
excited states with lifetimes on the order of tens of seconds or
longer~\cite{OMalley10}.
	
	\section*{Acknowledgments}
This work was supported by the Swedish Research Council (Contracts No.
821-2013-1642, No. 621-2015-04990, No. 621-2014-4501, and No. 621-2013-4084) and
by the Knut and Alice Wallenberg Foundation. We acknowledge support from the
COST Action No. CM1204 XUV/X-ray light and fast ions for ultrafast chemistry
(XLIC). M.K. acknowledges financial support from the Mobility Plus Program
(Project No. 1302/MOB/IV/2015/0) funded by the Polish Ministry of Science and
Higher Education. We thank J. Grumer and T. Brage (Lund University) for helpful
comments on theoretical aspects of the manuscript.
	
%
	
\end{document}